\pdfoutput=1
\documentclass[reprint,amsmath,amssymb,prl]{revtex4-1}
\usepackage{microtype}

\usepackage{hyperref}
\hypersetup{
    citecolor=blue,
    colorlinks=true,   
}

\begin{document}

\title{Comment on ``Poynting vector controversy in axion modified electrodynamics''}
\author{Kevin Zhou}
\email{knzhou@stanford.edu}
\affiliation{SLAC National Accelerator Laboratory, 2575 Sand Hill Road, Menlo Park, CA 94025, USA}

\begin{abstract}
In a recent publication, Tobar, McAllister, and Goryachev claim that using an alternative definition of the Poynting vector can dramatically enhance the sensitivity of their proposed axion dark matter detector. While the choice of Poynting vector is indeed ambiguous, it cannot affect physical results, such as the reading on a voltmeter, as long as the same definition is used consistently. I explain this point in detail, and locate the specific errors in their calculation. 
\end{abstract}

\maketitle

In the past decade, there has been great interest in the detection of ultralight axion dark matter. The axion may be described by a classical field $a(\mathbf{r}, t)$ with typical angular frequency $m_a$ and wavenumber $m_a v$, where $v \sim 10^{-3}$ is the dark matter virial velocity. In the presence of a static magnetic field $\mathbf{B}_0$, an electromagnetically coupled axion produces oscillating fields $\mathbf{E}_1$ and $\mathbf{B}_1$ proportional to the coupling $g$. If spatial gradients of the axion field are neglected, these electromagnetic fields can be found by adding an effective current $\mathbf{J}_{\text{eff}} = g \dot{a} \mathbf{B}_0$ to Ampere's law. 

When the length scale $R$ of the background field satisfies $m_a R \sim 1$, the resulting fields have typical size 
\begin{subequations}
\begin{align}
E_1 &\sim g a B_0, \label{eq:electric_res} \\
B_1 &\sim g a B_0.
\end{align}
\end{subequations}
They may be resonantly amplified in a microwave cavity, which is the basis for several ongoing experiments. More recently, experiments have been proposed for lighter axions in the ``quasistatic'' limit $m_a R \ll 1$, where~\cite{Ouellet:2018nfr}
\begin{subequations} \label{eq:standard_results}
\begin{align}
E_1 &\sim (g a B_0) (m_a R)^2, \label{eq:electric} \\
B_1 &\sim (g a B_0) (m_a R). \label{eq:magnetic}
\end{align}
\end{subequations}
Since the electric field is strongly suppressed, most experiments aim to amplify and detect $B_1$. However, Ref.~\cite{McAllister:2018ndu} proposed to detect the electric field, claiming that it would scale according to Eq.~\eqref{eq:electric_res}. This prompted several papers~\cite{Ouellet:2018nfr,Kim:2018sci,Beutter:2018xfx} which independently confirmed Eq.~\eqref{eq:electric}. 

In their latest paper~\cite{PhysRevD.105.045009}, the authors of Ref.~\cite{McAllister:2018ndu} argue that the sensitivity of their detector depends on the definition of the Poynting vector. Their core argument is preceded by a conceptual discussion, which is somewhat misleading. For instance, the authors claim that existing derivations of Eq.~\eqref{eq:standard_results} are incorrect because they neglect spatial gradients, setting $\nabla a = 0$. Spatial gradients are indeed important for some experiments, though in the quasistatic case, they only modify Eq.~\eqref{eq:standard_results} by terms subleading in $v$. But this issue is actually irrelevant to Ref.~\cite{PhysRevD.105.045009}, as the authors themselves set $\nabla a = 0$ above Eq.~(5.14) (i.e.~Eq.~(14) of Ref.~\cite{PhysRevD.105.045009}), before their main calculation. Many other ideas, such as duality symmetry, the Witten effect, magnetic currents, ferroelectrics, and antenna theory are mentioned throughout the text, but are also not actively used in their analysis.

Now let us review the Poynting vector $\mathbf{S} = \mathbf{E} \times \mathbf{B}$ in ordinary electromagnetism. In vacuum, a direct application of Maxwell's equations yields Poynting's theorem,
\begin{equation} \label{eq:poynting_theorem}
{-}\nabla \cdot \mathbf{S} = \frac{\partial}{\partial t} \left( \frac{E^2}{2} + \frac{B^2}{2} \right) + \mathbf{E} \cdot \mathbf{J}.
\end{equation}
The terms on the right-hand side are the rate of change of field energy and the rate of work done on matter, respectively. As a result, $\mathbf{S}$ is often verbally described as the flux of field energy, though one may add any divergence-free function to $\mathbf{S}$ and leave the form of Poynting's theorem unchanged~\cite{romer1982alternatives,gough1982poynting}. Alternatively, in dielectric and magnetic media, it may be convenient to use alternative forms of $\mathbf{S}$~\cite{kinsler2009four}. Each form may be useful in particular situations, but clearly, the choice cannot affect observable results, such as the reading on a voltmeter. For each choice of $\mathbf{S}$, there is an analogue of Poynting's theorem. But since this is merely a mathematical identity that is derived from Maxwell's equations, it doesn't contain any information that wasn't already present in the equations themselves. 

Similarly, the results of Ref.~\cite{PhysRevD.105.045009} are derived from Eqs.~(5.10) through (5.16), which I rewrite here for reference. The fixed, background static magnetic field satisfies
\begin{equation}
\nabla \times \mathbf{B}_0 = \mathbf{J}_0.
\end{equation}
This leads to order $g$ fields and sources, obeying
\begin{subequations}
\begin{align}
\nabla \cdot (\mathbf{E}_1 - g a \mathbf{B}_0) &= \rho_1 \\
\nabla \times \mathbf{B}_1 &= \partial_t (\mathbf{E}_1 - g a \mathbf{B}_0) + \mathbf{J}_1  \label{eq:modified_ampere} \\ 
\nabla \cdot \mathbf{B}_1 &= 0 \\
\nabla \times \mathbf{E}_1 + \partial_t \mathbf{B}_1 &= 0.
\end{align}
\end{subequations}
For clarity, I have used natural units and plugged in the authors' definitions of $\mathbf{D}_1$ and $\mathbf{H}_1$. The resulting equations are equivalent to the standard ones in the literature. 

The authors define an ``Abraham'' Poynting vector $\mathbf{S}_{\text{EH}}$ and a ``Minkowski'' Poynting vector $\mathbf{S}_{\text{DB}}$, which each describe the flow of power in the axion-induced field. I will focus on the latter, as the authors claim it leads to their anomalous results. Their derivation is written in terms of complex-valued fields, but for clarity I will use the equivalent real-valued fields, which are more familiar to physicists. First, Eq.~(5.24) is essentially
\begin{equation}
\mathbf{S}_{\text{DB}} = (\mathbf{E}_1 - g a \mathbf{B}_0) \times \mathbf{B}_1.
\end{equation}
Applying the above equations straightforwardly yields 
\begin{multline} \label{eq:poynting_alt}
{-}\nabla \cdot \mathbf{S}_{\text{DB}} = \frac{\partial}{\partial t} \left( \frac{E_1^2}{2} + \frac{B_1^2}{2} - g a \mathbf{B}_0 \cdot \mathbf{E}_1 + \frac{(gaB_0)^2}{2} \right) \\ + (- g a \mathbf{B}_0 \cdot \mathbf{J}_1 + g a \mathbf{B}_1 \cdot \mathbf{J}_0 + \mathbf{E}_1 \cdot \mathbf{J}_1)
\end{multline}
which is the differential form of Eq.~(5.28). 

Now consider the capacitor analyzed by the authors in section VI, with plates of radius $R$ and separation $d \ll R$, in the quasistatic limit $m_a R \ll 1$. Integrating both sides of Eq.~\eqref{eq:poynting_alt} over the strict interior $V$ of the capacitor, the terms involving the current vanish, leaving 
\begin{equation} \label{eq:integrated_poynting}
- \int_{\partial V} \mathbf{S}_{\text{DB}} \cdot d \mathbf{n} = \frac{\partial U_{\text{DB}}}{\partial t} 
\end{equation}
where the ``stored energy'' is 
\begin{equation} \label{eq:stored_energy}
U_{\text{DB}} = \int_V \frac{E_1^2}{2} + \frac{B_1^2}{2} - g a \mathbf{B}_0 \cdot \mathbf{E}_1 + \frac{(gaB_0)^2}{2}. 
\end{equation}
This is perfectly compatible with the standard results in Eq.~\eqref{eq:standard_results}, as can be straightforwardly checked, since both are derived from the same basic equations. 

In section VI.A.2, the authors perform a similar analysis but arrive at a different conclusion, due to several critical errors. First, their equivalent of Eq.~\eqref{eq:poynting_alt} is missing the $(gaB_0)^2/2$ term, even though it is much larger than the others~\footnote{Specifically, this term appears to be have been dropped without comment, below Eq.~(5.A4) in the appendix.}. Next, they neglect the ``insignificant'' $B_1^2/2$ term, even though it is one of the largest remaining terms~\footnote{In the text, this is justified by Eqs.~(5.44) and (5.45), which state that by Ampere's law, $\nabla \times \mathbf{B}_1 \approx \partial_t \mathbf{E}_1$, the magnetic field scales as $B_1 \sim (m_a R) E_1 \ll E_1$. This is incorrect because it ignores the $-\partial_t (g a \mathbf{B}_0)$ term in Eq.~\eqref{eq:modified_ampere}, which is much larger than $\partial_t \mathbf{E}_1$.}. Finally, they apply Eq.~(5.37), which was derived in the previous section for resonant cavities with $m_a R \sim 1$. It implies the scaling Eq.~\eqref{eq:electric_res}, so by assuming that it applies unchanged for a quasistatic haloscope, the authors have covertly assumed their conclusion. 

After these erroneous manipulations, the authors make a final, conceptual error. They equate $U_{\text{DB}}$ to 
\begin{equation}
U_c = C \mathcal{V}^2/2,
\end{equation} 
where $C$ is the capacitance and $\mathcal{V}$ is a ``voltage phasor'', concluding that a large measurable voltage is produced. However, the reading on a physical voltmeter is given by an integral over the wire path of the force per unit charge,
\begin{equation}
\mathcal{V} = \frac{1}{q} \int \mathbf{F} \cdot d \mathbf{s} = \int \mathbf{E}_1 \cdot d \mathbf{s}
\end{equation}
which only depends on $\mathbf{E}_1$, even in axion electrodynamics. Thus, this second definition of stored energy scales as $U_c \sim E_1^2 V$, while $U_{\text{DB}} \sim (gaB_0)^2 V$. But these are distinct definitions which cannot simply be equated~\footnote{Alternatively, if $\mathcal{V}$ is defined so that $U_{\text{DB}} = U_c$, it doesn't correspond to what a voltmeter measures, and hence cannot support claims of experimental sensitivity.}. 

A related conceptual error occurs in the authors' previous publications~\cite{Tobar:2018arx,Tobar:2020kmz}. Here, they rewrite the equations of axion electrodynamics in terms of a ``total'' electric field $\mathbf{E}^T = \mathbf{E} - g a \mathbf{B}$ (e.g.~see Eq.~(12.22)). Since the second term dominates over $E_1$ for quasistatic experiments, this redefinition automatically yields $E_1^T \sim gaB_0$. But this is irrelevant, because the reading on a voltmeter is still determined by $E_1$, not $E_1^T$. 

Stepping back, in the presence of background fields such as $\mathbf{B}_0$ and the axion field, there are multiple useful definitions of the Poynting vector, which correspond to multiple definitions of ``stored energy''. That is because while there is a definite total energy, it is ambiguous how to split it into a part due to the induced fields, and a part due to the background fields. It is precisely this ambiguity that lies at the heart of the Abraham--Minkowski controversy~\cite{griffiths2012resource}, which concerns different ways of splitting the total momentum of an electromagnetic wave in medium into contributions due to field and matter. 

But changing such definitions cannot change physical results unless algebraic errors are committed or inconsistent definitions are equated~\footnote{In fact, these errors also occur in their derivation using the ``Abraham'' Poynting vector, though there the errors happen to cancel each other, yielding the standard result.}, both of which occur in Ref.~\cite{PhysRevD.105.045009}. Thus, it cannot justify the claims of Ref.~\cite{McAllister:2018ndu}. 

\begin{acknowledgments}
I acknowledge support by the NSF GRFP under grant DGE-1656518. I thank Asher Berlin, Yonatan Kahn, Jonathan Ouellet, and Natalia Toro for useful discussions. 
\end{acknowledgments}

\bibliographystyle{apsrev4-1}
\bibliography{Comment}

\end{document}